\documentclass[5p]{article}

\usepackage{graphicx}
\usepackage{amssymb,amsthm}
\usepackage{color}

\begin{document}

\resizebox{!}{0.3cm}{\bf Alterations in brain connectivity due to plasticity}\\
\resizebox{!}{0.3cm}{\bf and synaptic delay}
\\
\\
E.L. Lameu$^1$, E.E.N. Macau$^{1,2}$, F.S. Borges$^3$, K.C. Iarosz$^{3}$, I.L. Caldas$^3$, R.R. Borges$^4$, P.R. Protachevicz$^5$, R.L. Viana$^6$, A.M. Batista$^{3,5,7}$
\\
$^1$National Institute for Space Research, S\~ao Jos\'e dos Campos, SP, Brazil.\\
$^2$Federal University of S\~ao Paulo, S\~ao Jos\'e dos Campos, SP Brazil.\\
$^3$Physics Institute, University of S\~ao Paulo, S\~ao Paulo, SP, Brazil.\\
$^4$Department of Mathematics, Federal Technological University of Paran\'a, Apucarana, PR, Brazil.\\
$^5$Science Post-Graduation, State University of Ponta Grossa, Ponta Grossa, PR, Brazil.\\
$^6$Physics Department, Federal University of Paran\'a, Curitiba, PR, Brazil.\\
$^7$Department of Mathematics and Statistics, State University of Ponta Grossa, Ponta Grossa, PR, Brazil.\\
\noindent Corresponding author: ewandson.ll@gmail.com

\date{\today}

\begin{abstract}
Brain plasticity refers to brain's ability to change neuronal
connections, as a result of environmental stimuli, new experiences, or damage.
In this work, we study the effects of the synaptic delay on both the coupling 
strengths and synchronisation in a neuronal network with synaptic plasticity. 
We build a network of Hodgkin-Huxley neurons, where the plasticity is given 
by the Hebbian rules. We verify that without time delay the excitatory synapses 
became stronger from the high frequency to low frequency neurons and the 
inhibitory synapses increases in the opposite way, when the delay is increased
the network presents a non-trivial topology. Regarding the synchronisation, only
for small values of the synaptic delay this phenomenon is observed.
\end{abstract}

keywords: magnetic surfaces, sympletic map, divertor



\section{Introduction}
Neuroplasticity, also known as brain plasticity, refers to brain's ability to
change neuronal connections, as a result of environmental stimuli, new
experiences, or damage \cite{leone05}. The brain plasticity can be functional
or structural. The functional plasticity occurs when functions are moved from
a damaged to other undamaged areas, and structural plasticity is associated with
changes in the physical structure \cite{kolb11}. On this regard, Borges et al.
\cite{borges16,borges17} studied the effects of the spike timing-dependent
plasticity (STDP) on the neuronal synchronisation. They observed that the
transition between desynchronised and synchronised states depends on the
external perturbation level and the neuronal architecture. It is know that neuronal
synchronisation is important in information binding \cite{lestienne01} and
cognitive functions \cite{wang10}. Nevertheless, synchronisation can be related
to brain disorders such as Parkinson's disease \cite{schwab13} and seizures
\cite{boucetta08}. This way, there have been many researches about not only
neuronal synchronisation \cite{protachevicz17}, but also suppression of
synchronous behaviour \cite{lameu16}.

We focus here on the effects of the synaptic delay on a neuronal
network with STDP. Information transmission delay is inherent due to both the
delays in synaptic transmission and the finite propagation velocities in the
conduction of signals \cite{kandel91}. Hao et al. \cite{hao11} studied
synchronisation transitions in a modified Hodgkin-Huxley neuronal network with
time delay. They found multiple synchronisation transitions when the time
delay is considered.

Experimental evidence of neuroplasticity was provide by Lashely in 1923
\cite{lashely23}. He dentified high  evidence of changes in neural pathways by means of
experiments on rhesus monkeys. More significant evidence began to be observed
in the 1960s. In 1964, Diamond et al. \cite{diamond64,bennett64} published
research about neuroplasticity, which is considered as the first evidence of
anatomical brain plasticity. Bach-y-Rita \cite{rita67} created a machine that
helped blind people not only to distinguish objects, but also to read. In 1949,
the neuropsychologist Donald Olding Hebb \cite{hebb49} wrote a book entitled
``The organization of behavior'', where he proposed that neurons which fire
together, also wire together. The Hebbian plasticity led model of spike
timing-dependent plasticity (STDP). The STDP function for excitatory and
inhibitory synapses were showed by Bi and Poo \cite{bi98} and Haas et al.
\cite{haas06}, respectively.

In this work, our results suggest that alterations in the synchronisation and
connectivity in a plastic network depend on the synaptic delay. We consider a
Hodgkin-Huxley neuronal network with inhibitory and excitatory neurons. The
Hodgkin-Huxley model \cite{hodgkin52} was proposed in 1952, and it is given by
coupled differential equations that explains the ionic mechanisms.

This paper is organised as follows: Section 2 introduces the Hodgkin-Huxley
neural network with synaptic delay. In Section 3, we introduce the synaptic
plasticity. In Section 4, we show our results about synaptic weights and
neuronal synchronisation. In the last Section, we draw the conclusions.


\section{Hodgkin-Huxley neural network with synaptic delay}

In the neuronal network we consider as local dynamics the neuron model proposed
by Hodgkin and Huxley in 1952 \cite{hodgkin52}. The individual dynamics of
each neuron in the network is given by
\begin{eqnarray}
C\dot{V_i} & = & I_i-g_{\rm K}n_i^{4}(V_i-E_{\rm K})-
g_{\rm Na}m_i^{3}h_i(V_i-E_{\rm Na}) -g_{L}(V_i-E_{\rm L})\nonumber \\
&+&\frac{(V_r^{\rm Exc}-V_i)}{\omega_{\rm Exc}}
\sum_{j=1}^{N_{\rm Exc}}\varepsilon_{ij} f_j(t) +\frac{(V_r^{\rm Inhib}-V_i)}{\omega_{\rm Inhib}}   \label{HH}
\sum_{j=1}^{N_{\rm Inhib}}\sigma_{ij} f_j(t),\\
\dot{n}_i & = & \alpha_{n_i}(V_i)(1-n_i)-\beta_{n_i}(V_i)n_i,\\
\dot{m}_i & = & \alpha_{m_i}(V_i)(1-m_i)-\beta_{m_i}(V_i)m_i,\\
\dot{h}_i & = & \alpha_{h_i}(V_i)(1-h_i)-\beta_{h_i}(V_i)h_i,
\end{eqnarray}
where $C$ ($\mu$F/cm$^2$) is the membrane capacitance and $V_i$ (mV) is the
membrane potential of neuron $i$ ($i=1,...,N$). $I_i$ represents a constant
current density that is randomly distributed in the interval $[9.0;10.0]$,
$\omega_{\rm Exc}$ (excitatory) and $\omega_{\rm Inhib}$ (inhibitory) are the
average degree connectivities, $\varepsilon_{ij}$ and $\sigma_{ij}$ are the
excitatory and inhibitory coupling strengths from the presynaptic neuron $j$ to
the postsynaptic neuron $i$. $N_{\rm Exc}$ and $N_{\rm Inhib}$ are the number of
excitatory and inhibitory neurons, respectively. The parameters $g_K, g_{Na}$
and $g_L$ are the condutances of the potassium, sodium and leak ion channels,
respectively. $E_K, E_{Na}$ and $E_L$ are the reversal potentials for these ion
channels. The functions $m(V_i)$ and $n(V_i)$ represent the activation for
sodium and potassium, respectively. $h(V_i)$ is the function for the
inactivation of sodium. The functions $\alpha_{n}$, $\beta_{n}$, $\alpha_{m}$,
$\beta_{m}$,$\alpha_{h}$, $\beta_{n}$ are given by
\begin{eqnarray}
\alpha_{n}(v) & = & \frac{0.01 v + 0.55}{1 - \exp \left(-0.1 v-5.5 \right)},\\
\beta_{n}(v) & = & 0.125\exp\left(\frac{-v-65}{80}\right),\\
\alpha_{m}(v) & = & \frac{0.1 v + 4}{1 - \exp\left (-0.1 v - 4\right)},\\
\beta_{m}(v) & = & 4\exp\left(\frac{-v-65}{18}\right),\\
\alpha_{h}(v) & = & 0.07\exp\left(\frac{-v-65}{20}\right),\\
\beta_{h}(v) & = & \frac{1}{1 + \exp\left(-0.1 v - 3.5\right)},
\end{eqnarray}
where $v=V/[mV]$. The neuron can present periodic spikings or single
spike activity as a result of the variation of the external current density
$I_i$ ($\mu$A/cm$^2$). The frequency of the periodic spikes increases if the
constant $I_i$ increases. 

\begin{figure}[htbp]
\begin{center}
\includegraphics[height=6cm,width=10cm]{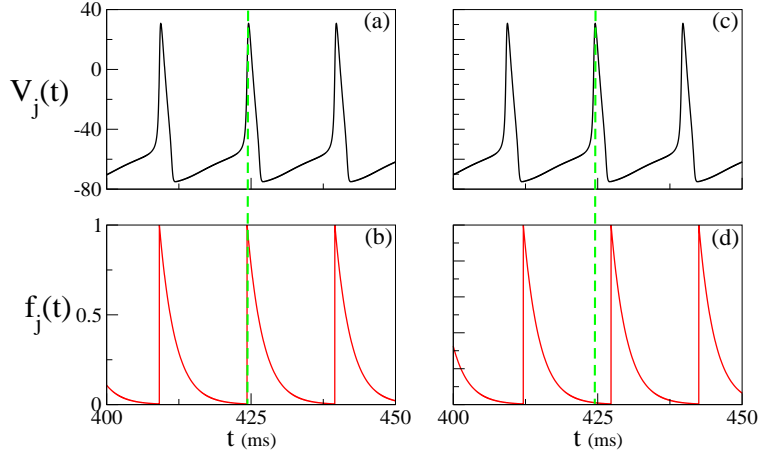}
\caption{Time evolution of the action potential $V_j(t)$ of a presynaptic neuron
$j$ and the respective synaptic current (output) $f_j(t)$ that achieves the
postsynaptic neuron $i$. We consider $\tau=0.0$ in (a) and (b), and
$\tau = 3.0$ ms in (c) and (d).}
\label{fig1}
\end{center}
\end{figure}

In Equation (\ref{HH}) the term $f_j(t)$ is a function which represents the
strength of an effective synaptic (output) current and it is given by
\begin{eqnarray}
f_j(t) & = & e^\frac{-(t-t_j-\tau)}{\tau_s}, \label{s}
\end{eqnarray}
where $\tau_s$ is the synaptic time constant and $t_j$ is the most recent
firing instant of the neuron $j$. The parameter $\tau$ is the time delay and
consequently the time that the current $f_j(t)$ spends to achieve the
postsynaptic neuron \cite{hao11}. Figures \ref{fig1}(a) and \ref{fig1}(c) show
the time evolution of the action potential $V_j(t)$ for $\tau=0$ and
$\tau=3$ ms, respectively. The action potential starts at $-70$ mV and when a
stimulus is applied it spikes upward. After the peak potential, the action
potential falls to the resting potential. In Figures \ref{fig1}(b) and
\ref{fig1}(d) we calculate $f_j(t)$ for the respective Figures \ref{fig1}(a)
and \ref{fig1}(c). We see by means of the dashed green line that the
transmission of the synaptic current to the postsynaptic is not instantaneous
for $\tau=3$ ms.

In our simulations, we consider $C=1$ $\mu$ F/cm$^{2}$, $E_{\rm Na}=50$ mV,
$E_{\rm K}=-77$ mV, $E_{\rm L}=-54.4$ mV, $g_{\rm Na}=120$ mS/cm$^{2}$,
$g_{\rm K}=36$ mS/cm$^{2}$, $g_{\rm L}=0.3$ mS/cm$^{2}$ and $\tau_s=2.728$ ms. The
neurons are excitatorily coupled with a reversal potential $V^{\rm Exc}_{r}=20$ mV,
and inhibitorily coupled with a reversal potential $V^{\rm Inhib}_{r}=-75$ mV
\cite{borges17}.

\section{Synaptic plasticity}

Synaptic plasticity is the process that produces changes in the synaptic
strength, namely it is the strengthening or weakening of synapses over time.
In 1998, the neuroscientists Bi and Poo \cite{bi98} characterised the dependence
of the long-term potentiation and depression on the order and timing of pre and
postsynaptic spikes, named spike time dependant plasticity (STDP). The
plasticity dynamics is given by the update value of the synaptic weight
$\Delta\Gamma$, and a mathematical definition of this function is given by
\cite{kalitzin00}
\begin{equation}\label{edo}
{d\Delta\Gamma(t) \over dt}=y(\Delta\Gamma,V,t).
\end{equation}
Kalitzin and collaborators \cite{kalitzin00} showed that the function $y$ depends on the
membrane potential of the postsynaptic neuron, the activation of the synapse,
and the thresholds for switching on long-term potentiation and the long-term
depression. We consider an approximation of $y$ in the linear form 
$y(\Delta\Gamma,t)=(a+c/t)\Delta\Gamma$ \cite{borges17}. The function 
$\Delta\Gamma=bt^c\exp(at)$ is the solution of Equation (\ref{edo}), where
$a$, $b$, and $c$ are constants. For $c=0$, we obtain the update value for
excitatory synapses $\Delta\varepsilon$ (eSTDP), and for $c\neq 0$, we find the update value for
inhibitory synapses $\Delta\sigma$ (iSTDP). The plasticity dynamics introduced by means of
this linear approximation is not related to physiological processes
\cite{artola90}, however, with this function we can find a fit which describes
experimental results of eSTDP and iSTDP, as showed in References \cite{bi98} and
\cite{haas06}. 

\begin{figure}[htbp]
\begin{center}
\includegraphics[height=7cm,width=8cm]{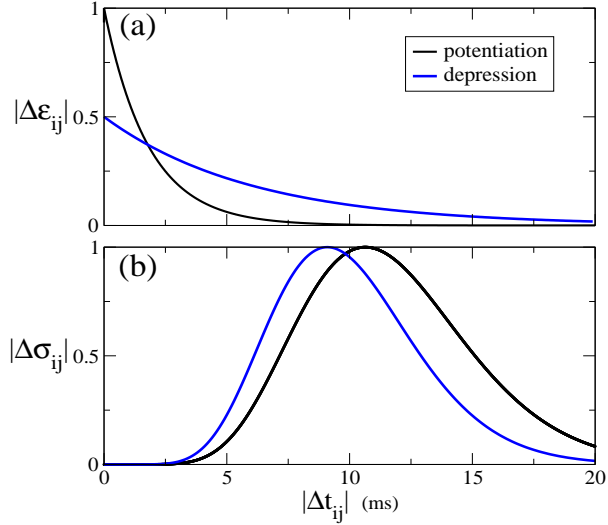}
\caption{Comparison between absolute values of potentiation (black curves)
versus depression (blue curves) in synaptic weights. STDP function for (a)
excitatory and (b) inhibitory synapses.}
\label{fig2}
\end{center}
\end{figure}

Figure \ref{fig2}(a) exhibits the eSTDP function for excitatory synapses,
where the presynaptic neuron $j$ and the postsynaptic neuron $i$ are forced
to spike at time $t_{j}$ and $t_{i}$, respectively. There is a change in the
synaptic weights $\Delta \varepsilon_{ij}$ due to the time difference between
the spikes $\Delta t_{ij}=t_{i}-t_{j}$. 
The eSTDP function is given by \cite{bi01}
\begin{equation}\label{eqplast}
\Delta \varepsilon_{ij}= \left\{
\begin{array}{ll}
\displaystyle A_{1}\exp(-\Delta t_{ij}/\tau_{1}),\;\; if \;\;\Delta t_{ij}\geq 0 \\
\displaystyle -A_{2}\exp({\Delta t_{ij}/\tau_{2}}),\;\;if \;\; \Delta t_{ij} < 0
\end{array}
\right. ,
\end{equation}
where $A_{1}=1$, $A_{2}=0.5$, $\tau_{1}=1.8$ms, and $\tau_{2}=6$ms. The synaptic
weights are updated according to Equation (\ref{eqplast}), where 
$\varepsilon_{ij}\rightarrow \varepsilon_{ij}+10^{-3}\Delta\varepsilon_{ij}$. 
The black line in figure \ref{fig2}(a) shows the potentiation of 
excitatory synaptic weights for $\Delta t_{ij}\geq 0$ and the blue line the 
depression in synaptic weights for $\Delta t_{ij} < 0$.

In Figure \ref{fig2}(b), we see the iSTDP function for inhibitory synapses. 
The weights are increased based on the following equation
\begin{equation}\label{eqplastI}
\Delta \sigma_{ij} =  \frac{g_0}{g_{\rm norm}} {\alpha}^{\beta} |\Delta t_{ij}| 
{\Delta t_{ij}}^{\beta -1} \exp(-\alpha |\Delta t_{ij}|),
\end{equation}
where $g_0 = 0.02$, $\beta=10$, $\alpha = 0.94$ if $\Delta t_{ij}>0$, 
$\alpha=1.1$ if $\Delta t_{ij}<0$ and $g_{\rm norm} = {\beta}^{\beta}  \exp(-\beta)$ 
\cite{talathi08,abarbanel06}. The inhibitory synaptic weights are updated
according to Equation (\ref{eqplastI}), where
$\sigma_{ij}\rightarrow \sigma_{ij}+10^{-3}\Delta\sigma_{ij}$. 

In our neural network model, the time interval between spikes $\Delta t_{ij}$ and 
the plasticity rules are calculated and applied every time the postsynaptic neuron 
$i$ fires and can present different values
depending on when the presynaptic neuron $j$ had the last spike.


\section{Synaptic weights and synchronisation}

\begin{figure}[htbp]
\begin{center}
\includegraphics[height=9cm,width=10cm]{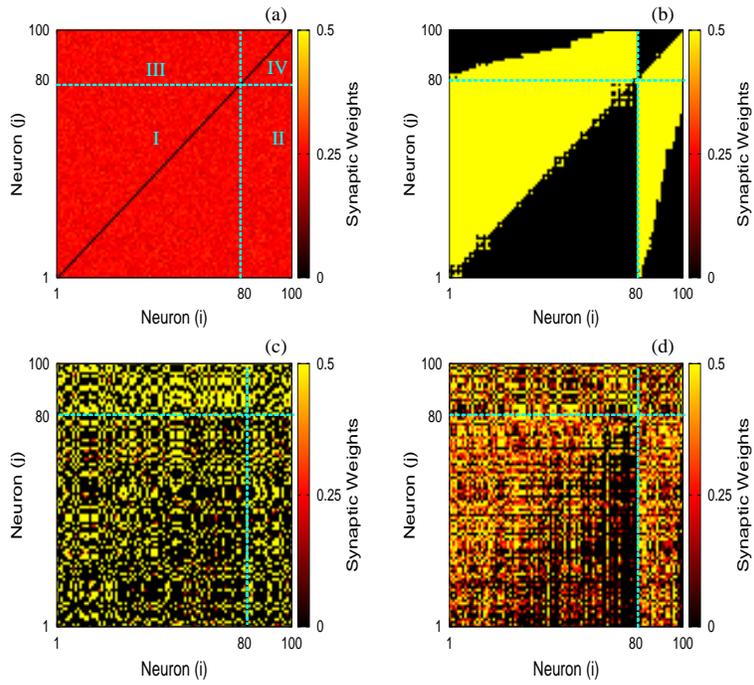}
\caption{Coupling matrices for excitatory and inhibitory neurons. Figure (a)
shows the initial synaptic weights. We consider (b) $\tau=0$ ms, (c) $\tau=3$ ms
and (d) $\tau=6$ ms at $400$ s. In four cases the colour bar represents the
synaptics weights.}
\label{fig3}
\end{center}
\end{figure}

In our simulations, aiming to understand the alterations in network
connectivity, we consider a neuronal network with $100$ Hodgkin-Huxley. This
number of neurons was chosen to facilitate a visual analysis of the coupling
matrices without to lose dynamics properties. Our network has $80\%$ of 
excitatory and $20\%$ of inhibitory synapses according to anatomical estimates
for the neocortex \cite{noback}. The neurons are initially globally coupled and
the initial synaptic weights are normally distributed with mean $0.25$ and
standard deviation equal to $0.02$. In this approach, to understand 
the impact of the delay in the system, we will consider that all the synapses 
have the same delay. In Figure \ref{fig3} we see the coupling
matrices, where the colour bar represents the synaptic weights. The coupling
matrix is separated into excitatory ($1\leq i,j\leq 80$) and inhibitory
($81\leq i,j \leq 100$) neurons. The excitatory neurons $i$ are organised from
the lowest frequency $i=1$ to the highest frequency $i=80$, and the inhibitory
neurons from the lowest frequency $i=81$ to the highest frequency $i=100$.

Figure \ref{fig3}(a) exhibits the initial synaptic weights separated into $4$
regions. In the regions I and II the synapses from the pre to the postsynaptic
neurons are excitatory. The region III and IV have inhibitory synapses from the
pre to postsynaptic neurons. For $\tau=0$ms, we observe in Figure \ref{fig3}(b)
that the coupling matrix shows a triangular shape, due to the fact that the
excitatory synapses become stronger from the high to low frequency neurons
and the inhibitory synapses from the low to high frequency neurons.
When the time delay is $\tau=3$ms and also $\tau=6$ms, as shown in Figures
\ref{fig3}(c) and \ref{fig3}(d), respectively, the coupling matrices have a
non-trivial configuration of connections, presenting a greater agreement with 
real neuronal networks \cite{sarah,yu,bonifazi}. Therefore, the time delay has
a significant influence on the synaptic weights in a neuronal network with
plasticity, resulting in non-trivial configurations and synaptic weights 
with greater variability in their values if compared to the case without delay

\begin{figure}[htbp]
\begin{center}
\includegraphics[height=8cm,width=12cm]{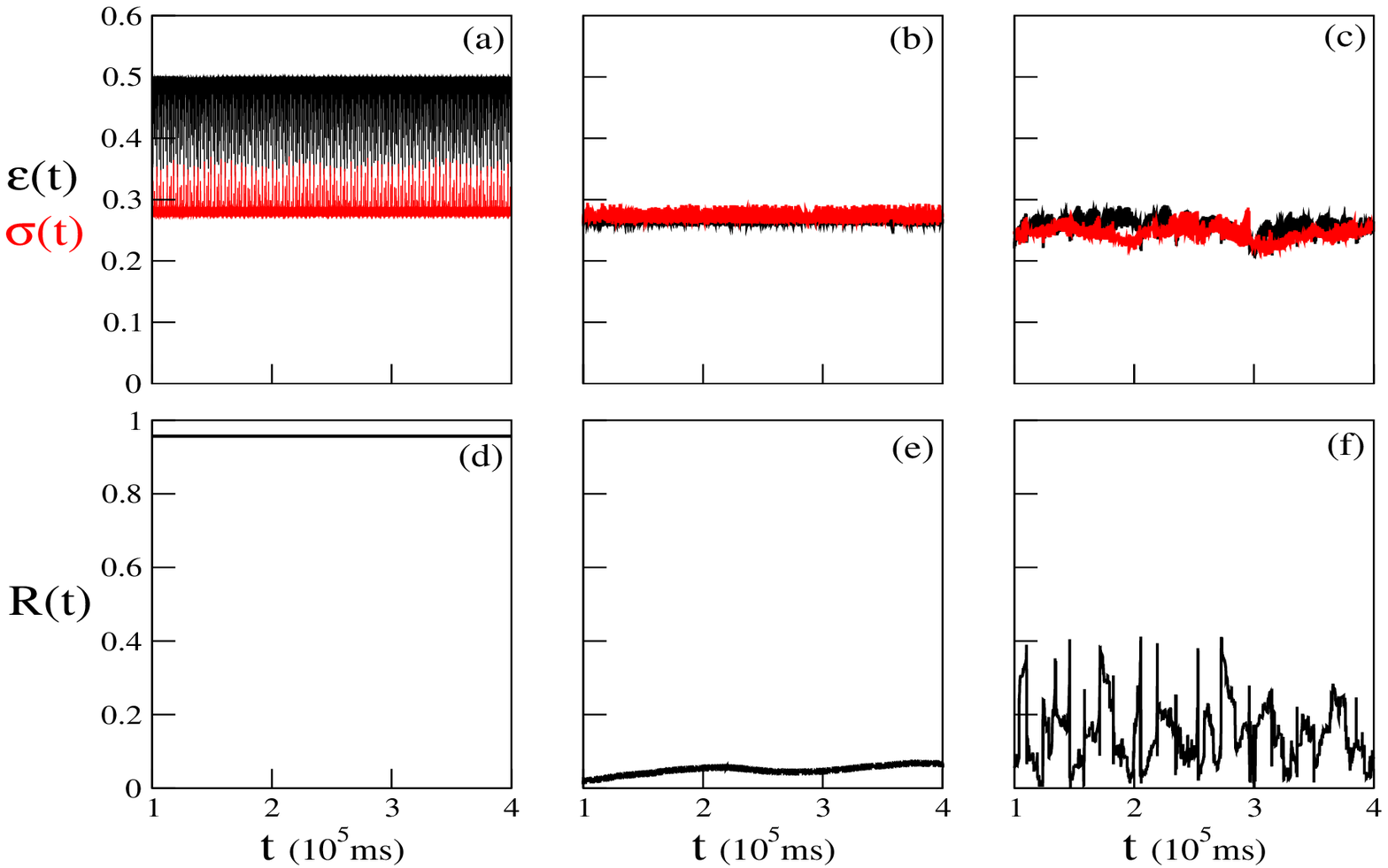}
\caption{In this set of Figures we show the time evolution of $\varepsilon(t)$
(black line) and $\sigma(t)$ (red line) for (a) $\tau=0$, (b) $\tau=3$ ms, and
(c) $\tau=6$ ms, as well as the time evolution of the Kuramoto's order
parameter $R(t)$ for(d) $\tau=0$, (e) $\tau=3$ ms, and (f) $\tau=6$ ms.}
\label{fig4}
\end{center}
\end{figure}

We analyse the time evolution of instantaneous average of excitatory
$\varepsilon(t)$ and $\sigma(t)$ inhibitory coupling strengths for different
time delay values. Without time delay $\tau=0$ (Figure \ref{fig4}(a)),
$\varepsilon$ (black line) has value greater than $\sigma$ (red line). Whereas
for $\tau=3$ms (Figure \ref{fig4}(b)) and $\tau=6$ms (Figure \ref{fig4}(c)) both
$\varepsilon$ and $\sigma$ oscillate in the interval $[0.2;0.3]$.

We study the effects of the time delay on the neuronal synchronisation. To do
that, we use the Kuramoto order parameter as diagnostic tool, that is given by
\cite{kuramoto84}
\begin{equation}
R(t)= \left| \frac{1}{N}\sum_{j=1}^{N}\exp(i\phi_{j}(t))\right| , \label{Rtempo}
\end{equation} 
and the time averaged order parameter
\begin{equation}
{\bar R}=\frac{1}{t_{\rm f}-{t_{\rm i}}}\sum_{t_{\rm i}}^{t_{\rm f}}
\left| \frac{1}{N}\sum_{j=1}^{N}\exp(i\phi_{j}(t))\right| , \label{Rmedio}
\end{equation} 
where $\phi_{j}(t)$ is the phase associated with the spikes,
\begin{equation}
\phi_{j}(t)=2\pi m+2\pi\frac{t-t_{j,m}}{t_{j,m+1}-t_{j,m}},
\end{equation}
where $t_{\rm f}-t_{\rm i}$ is the time windows for measuring, $t_{j,m}$ is the
time when a spike $m$ ($m=0,1,2,\dots$) in the neuron $j$ happens
($t_{j,m}<t<t_{j,m+1}$). The order parameter magnitude asymptotes to unity when
the network has a globally synchronised behaviour. For uncorrelated spiking
phases, the order parameter is much less than 1.

Figures \ref{fig4}(d), \ref{fig4}(e), and \ref{fig4}(f) exhibit the order
parameter for (d) $\tau=0$, (e) $\tau=3$ms, and (f) $\tau=6$ms. Our neuronal
network does not exhibit completely synchronisation due to the fact that the
neurons are not identical. Nevertheless, for $R>0.9$ the neuronal network shows
strong synchronisation behaviour. In Figure \ref{fig4}(d), we see a synchronous
state for $\tau=0$. There is no synchronisation states observed for $\tau=3$ms
and $\tau=6$ms, as shown in Figures \ref{fig4}(e) and \ref{fig4}(f),
respectively. This result shows that the delay is an important mechanism in the
network dynamics, avoiding synchronization.

\begin{figure}[htbp]
\begin{center}
\includegraphics[height=8cm,width=7cm]{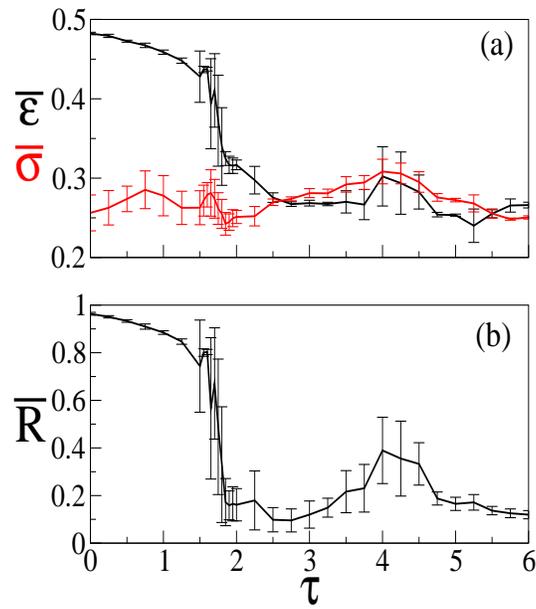}
\caption{(a) Average values of excitatory ${\bar \varepsilon}$, inhibitory
${\bar \sigma}$ synaptic weights, and (b) mean order parameter ${\bar R}$ as a
function of synaptic time delay $\tau$. The bars show the standard deviation
from the mean values.}
\label{fig5}
\end{center}
\end{figure}

In Figures \ref{fig5}(a) we calculate the time averaged excitatory and
inhibitory coupling strengths as a function of the time delay for $10$
different initial conditions.
The ${\bar \sigma}$ values presents a small variation as the delay $\tau$ is 
increased. However, ${\bar \varepsilon}$ is more sensitive and for small 
delay values $\tau<1.5$ms we observe ${\bar \varepsilon}>{\bar \sigma}$ 
and the network is more excitable. As a result the neurons in the network 
are strongly synchronized (Figure  \ref{fig5}(b)). When we increase the 
delay for $\tau>1.5$ ms the values of ${\bar \varepsilon}$ starts to 
decrease in a second order transition. Simultaneously the order parameter  
${\bar R}$ decreases showing its dependence with the excitatory coupling 
strength ${\bar \varepsilon}$. Finally, for  $\tau>2.5$ ms we observe that 
${\bar \varepsilon}$ and ${\bar \sigma}$ oscillates in the interval $[0.2;0.3]$ 
and the network are no longer synchronized. These results show us that 
synchronization in a neuronal network with plasticity and synaptic delay 
is closely linked to the intensity of excitatory couplings, i.e, the more 
excitable the network (${\bar \varepsilon}>{\bar \sigma}$) the more synchronous
the neurons will be.


\section{Conclusion}

We study a neural network with plasticity and synaptic delay, where we
consider the Hodgkin-Huxley model as local dynamics. The Hodgkin-Huxley neuron
is a mathematical model described by coupled differential equations that
exhibits spiking dynamics. We build a network with an initial all-to-all
topology and analyse the time evolution of the connectivity and synchronisation.

We carry out simulations considering a coupling matrix with initial synaptic
weights normally distributed. Without time delay, the coupling matrix evolves
to a triangular shape, where the excitatory synapses are stronger from the high 
frequency to low frequency neurons an the inhibitory synapses increases in the
opposite way. The coupling matrix exhibits non-trivial configuration
when the time delay is increased.

We also show that the time delay plays an important role in the neural
synchronisation. Increasing the time delay, we verify that the time averaged
excitatory coupling strength decrease and it becomes approximately equal to
the averaged inhibitory coupling strength. As a consequence, this decrease
suppresses the synchronous behaviour of the neural network.


\section*{Acknowledgments}
This work was possible by partial financial support from the following
Brazilian government agencies: CNPq (154705/2016-0, 311467/2014-8), CAPES,
Funda\-\c c\~ao Arau\-c\'aria, and S\~ao Paulo Research Foundation (processes
FAPESP 2011/19296-1, 2015/ 07311-7, 2016/23398-8, 2015/50122-0).
Research supported by grant 2015/50122-0 Sao Paulo Research Foundation (FAPESP)
and DFG-IRTG 1740/2.


\begin{frame}

\end{frame}
\end{document}